\documentclass[fleqn,10pt]{wlscirep}

\usepackage{xcolor}
\usepackage{bm}

\title{Echo Chambers: Emotional Contagion and Group Polarization on Facebook}
\author[1]{Michela Del Vicario}
\author[1]{Gianna Vivaldo}
\author[1,2]{Alessandro Bessi}
\author[1]{Fabiana Zollo}
\author[1,3]{Antonio Scala}
\author[1]{Guido Caldarelli}
\author[1,*]{Walter Quattrociocchi}
\affil[1]{Laboratory of Computational Social Science, Networks Dept, IMT School for Advanced Studies, 55100 Lucca, Italy}
\affil[2]{IUSS Institute for Advanced Study, Piazza della Vittoria 5, 27100 Pavia, Italy}
\affil[3]{ISC-CNR Uos "Sapienza", 00185 Roma, Italy}
\affil[*]{walter.quattrociocchi@gmail.com}


\begin{abstract}
Recent findings showed that users on Facebook tend to select information that adhere to their system of beliefs and to form polarized groups -- i.e., echo chambers. 
Such a tendency dominates information cascades and might affect public debates on social relevant issues.
In this work we explore the structural evolution of communities of interest by accounting for users emotions and engagement. 
Focusing on the Facebook pages reporting on scientific and conspiracy content, we characterize the evolution of the size of the two communities by fitting daily resolution data with three growth models -- i.e. the Gompertz model, the Logistic model, and the Log-logistic model. 
Then, we explore the interplay between emotional state and engagement of users in the group dynamics.
Our findings show that communities' emotional behavior is affected by the users' involvement inside the echo chamber. Indeed, to an higher involvement corresponds a more negative approach. Moreover, we observe that, on average, more active users show a faster shift towards the negativity than less active ones.
\end{abstract}

\begin{document}
	
\flushbottom
\maketitle
\thispagestyle{empty}
	
\section*{Introduction}
	
Misinformation has traditionally represented a political, social, and economic risk. The digital age, in which new ways of communication arose, has exacerbated its extent, and mitigation strategies  are even more uncertain.
However, according to the World Economic Forum, massive digital misinformation remains one of the main threats to our society\cite{WEF16}.

The diffusion of social media caused a shift of paradigm in the creation and consumption of information. We passed from a mediated (e.g., by journalists) to a more disintermediated selection process. Such a disintermediation elicits the tendencies of the users to a) select information adhering to their system of beliefs -- i.e., confirmation bias -- and b) to form groups of like-minded people where they polarize their opinion -- i.e. echo chamber\cite{Cacciatore2015,brown2007word, Richard2004, QuattrociocchiCL11,Quattrociocchi2014,Kumar2010}.
	
Under these settings, discussion within like-minded people seems to negatively influence users' emotions and to enforce group polarization\cite{sunstein2002law, zollo2015emotional}. What's more, experimental evidence shows that confirmatory information gets accepted even if containing deliberately false claims\cite{bessi2015science, del2015spreading, bessi2014economy, bessi2015viral, bessi2016homophily}, while dissenting information is mainly ignored or might even increase group polarization\cite{zollo2015debunking}.
Current solutions, such as debunking efforts or algorithmic driven solutions based on the reputation of the source, seem to be ineffective \cite{ciampaglia2015computational,qazvinian2011rumor}. To make things more complicated, users on social media aim at maximizing the number of likes (Attention Bulimia) and often information, concepts, and debate get flattened and oversimplified.
In such a disintermediated environment, indeed, the public opinion deals with a large amount of misleading information that might influence important decisions.
	
Computational social science\cite{Lazer721} seems to be a powerful tool for a better understanding of the cognitive and social dynamics behind misinformation spreading\cite{WEF16}.	
Along this path, in the present work we address the evolution of online echo chambers by performing a comparative analysis  of two distinct polarized communities on the Italian Facebook, i.e., science and conspiracy. 
The sizes of both the communities are firstly analyzed in terms of their temporal evolution and fitted by classical population growth models deriving from biology and medicine fields. The behavior of users turns out to be similar for both categories, irrespective of the contents: both science and conspiracy communities reach a thresholding value in their sizes, after an almost exponential growth, in agreement with classical growth models.

Moreover, we analyze the community behavior by accounting for the engagement and the emotional dynamics of users.
Indeed, whether a news item, either substantiated or not, is accepted as true by a user may be strongly affected by social norms or by how much it coheres with the community shared system of beliefs. 
	
Users' emotional behavior seems to be affected by their engagement within the community. An higher involvement in the echo chamber, resolves in a more negative emotional state. Such a phenomenon appears in both users categories.  Moreover, we observe that, on average, more active users show a faster shift towards the negativity than less active ones. 
	
The paper is structured as it follows. First we analyze the structural evolution of both science and conspiracy communities on the Italian Facebook. Then we explore the user sentiment behavior as a single unit, and subsequently we explore the sentiment contagion inside each of the two communities from a macroscopic point of view.

\section*{Results and Discussion}
\subsection*{Community Evolution}
Online social networks might elicit the aggregation of individuals in communities of interest. 
For the particular case of science and conspiracy users on the Italian Facebook (refer to section \textit{Methods} for more details on the data collection and classification), the emergence of two separate echo chambers has already been shown in a previous study\cite{del2015spreading}. 
However little is known about the structural evolution of the two communities and the role of users' engagement in shaping them.
To shade light on the determinants of group formation, as a first step, we analyze and compare the temporal evolution of science and conspiracy communities size by considering users commenting activity.
	
More in details, we divide users in three categories : 
	
\begin{itemize}
	\item $U_1$ the set of all active users -- i.e. of all those users that commented at least once,
	\item $U_2$ the set of all users that commented at least twice, and
	\item $U_5$ the set of all users that commented at least five times. 
\end{itemize}
For each set of users we look at the temporal evolution of the science and conspiracy communities, defined as:
$$
S_i(t) = \left\{ u \in U_i: \frac{s_u}{n_u} \geq 0.95 \right\}\mbox{ and }C_i(t) = \left\{ u \in U_i: \frac{c_u}{n_u} \geq 0.95 \right\},
$$
	
where $i\in\{1,2,5\}$, $n_u$ is the total number of comments made by user $u$, $s_u$ is the number of comments that user $u$ made on science posts, $c_u$ is the number of comments that user $u$ made on conspiracy posts, and  $t\in\{1,\ldots, T\}$\footnote{The observation is carried out over the period January 2010--April 2012, by daily temporal	resolution. $T$ is the number of days of observation and it is equal to $835$.}. We consider the threshold of $0.95$ for the membership inside one community in accordance with previous studies \cite{bessi2015science, bessi2014economy}.
	
\begin{figure}[h]
	\centering
    \includegraphics[width=1\textwidth]{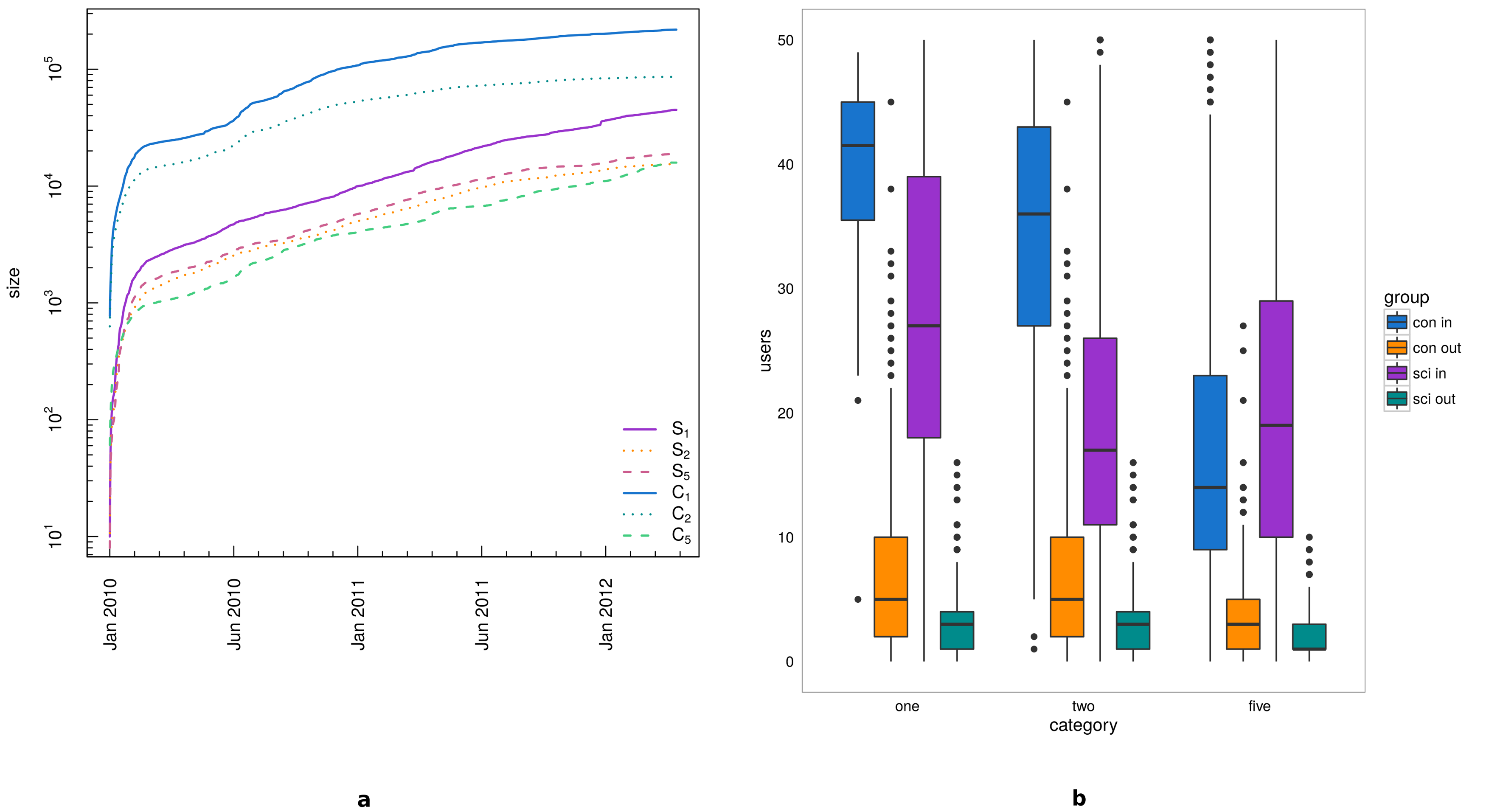}
	\caption{(a) Temporal evolution of the size of the communities $S_1$ (solid violet), $S_2$ (dotted orange), $S_5$ (dashed pink), $C_1$ (solid blue), $C_2$ (dotted sea green), and $C_5$ (dashed green). The observation is carried out in the period January 2010--April 2012, with daily temporal resolution. (b) Boxplots of the users' mobility within each group. The black horizontal lines represent the median of the number of users entering or exiting the science and conspiracy communities for each temporal step. Individual points refer to the outliers of the distributions. Colored boxes represent the interquartile range (25th–75th percentile range), where blue stands for incoming users in conspiracy community, violet for incoming users in science, orange for exiting users from conspiracy, and green for exiting users from science. On the x-axis we have, from left to right, results for $C_1$, $S_1$, $C_2$, $S_2$, $C_5$, and $S_5$. }\label{fig:community_evolution}
\end{figure}
	
Figure~\ref{fig:community_evolution}(a) shows the temporal evolution of the size of the communities resulting from the previous classification. The dataset has been sampled by daily resolution, over the period January 2010--April 2012, for a total of 835-days observations. A similar global behavior emerges in all cases, and significant quantitative differences arise between $C_1$ (or $C_2$) and $C_5$, as well as between $C_1$ (or $C_2$) and $S_i$, $i \in \{1,2,5\}$. This phenomenon may be linked to the abundance of low-activity users inside the conspiracy communities, and for this reason in the next sections we will restrict our attention to the respective most active communities, $S_5$ and $C_5$.
We also pairwise compared the six sample distributions by means of the Kolmogorov-Smirnov test (see Tab.~\ref{tab1} for the tests' results). For each users typology, we reject the null hypothesis of equivalence between science and conspiracy distributions, at the $99\%$ confidence level.
\begin{table}[ht]
	\centering
	\begin{tabular}{c|c|c|c}
			
		& $D$&  $C$ & $p$ \\ 
		
		$S_1/C_1$ & $0.763$ & $0.079$ & $2.2\times10^{-16}$ \\
		$S_2/C_2$ & $0.886$ & $0.079$ & $2.2\times10^{-16}$ \\
		$S_5/C_5$ & $0.970$ & $0.079$ & $2.2\times10^{-16}$ \\
	\end{tabular}
	\caption{Results from Kolmogorov-Smirnov tests. $D$ is the estimated maximum distance between the two distributions under analysis, $C$ is the corresponding critical value, and $p$ the resulting p-value. Considering a level of significance $\alpha = 0.01$, we reject the null hypothesis of equivalence of the two distributions in all the cases. }\label{tab1}
\end{table}

In Fig.~\ref{fig:community_evolution}(b) we report the summarizing statistics for the users' mobility inside one particular community by box and whiskers plots\cite{chambers1983graphical} (or, simply boxplots). Black horizontal lines represent the median of the number of users entering or exiting the science and conspiracy communities, and the colored boxes represent the interquartile ranges (i.e., the $25$th–$75$th percentile ranges) and they statistically measure the degree of dispersion and the skewness of each analyzed distribution: the users which enter the science and conspiracy communities (violet and blue boxes, respectively), and the users which exit from each community (green and orange boxes for science and conspiracy, respectively). Vertical lines (i.e., the whiskers) are lower and upper bounded by the minimum and maximum values of the corresponding distribution, once both outliers and extreme values are removed from the data. Individual points represent the outliers of each analyzed distribution. From the left to the right, each set of boxplots corresponds to one user's category (i.e., $U_1$, $U_2$, and $U_5$). 
\\
In all cases we notice a significant difference between the users entering into and exiting from a community, favorable to the formers, indeed more than $99\%$ of the users' flow is made up of those users entering one community.
\\
These two results underline that the behavior of users is similar for both categories, irrespective of the contents. After an initial spike-like growth, the communities evolve at a nearly constant rate. Moreover, once a user enters one community the probability to get out of it is very small.

To better characterize the temporal evolution of both communities, we fit the Gompertz growth model (GM) in (\ref{GM}), the Logistic model (LM$3$, LM$5$) in (\ref{LM}), and the Log-logistic model (LLM) in (\ref{LLM}) to our sample distributions $S_5$ and $C_5$, representing the temporal profile of quite active users, i.e. with at least 5 total comments, affiliated to science or conspiracy communities, respectively. The models are chosen on the basis of the observed evolution of the communities' size, that is characterized by a first phase of rapid growth, approximately exponential, followed by a more gradual one.

\begin{figure}[h]
	\centering
	\includegraphics[width=1\textwidth]{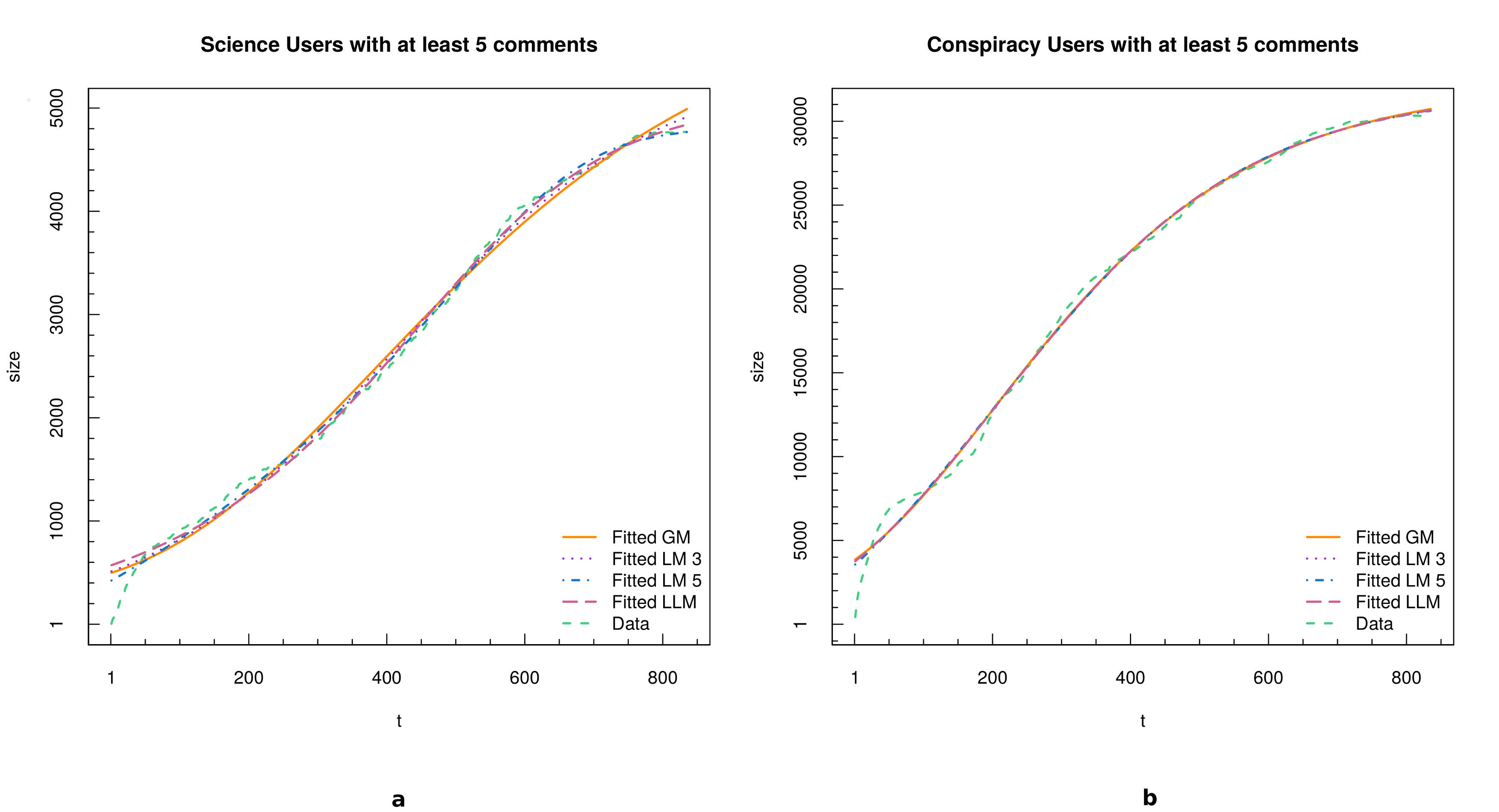} 
	\caption{Fit of the temporal evolution of the size of science (a) and conspiracy (b) communities. We fitted the data with four growth models: GM (bold orange line), LM$3$ (dotted violet line), LM$5$ (dashed-dotted blue line), and LLM (dashed purple line). All models show a good fit for our data samples. The temporal scale is represented by daily temporal steps, starting from January 2010 ($t=0$) till April 2012 ($t=835$).}\label{fig:community_fit}
\end{figure}

For each model we estimate its parameters through the \textit{Nonlinear Least Squares} \textit{NLS} (see Section~\textit{Methods} for more details about the fitting models). 
Fit's results are shown in Fig.~\ref{fig:community_fit} for both science (panel a) and conspiracy (panel b). Four fits are superposed to original data (green dotted line): GM (bold orange line), LM$3$ (dotted violet line), LM$5$ (dashed-dotted blue line), and LLM (dashed purple line).
	
As it can be deduced from Fig.~\ref{fig:community_fit}, all models show a good approximation of the temporal evolution of science and conspiracy communities sizes. Anyway, in order to identify the best fit and the quality of each fit, we perform a series of Kolmogorov-Smirnov tests (KS) between the real data and each of the synthetic distributions and Maximum Likelihood Estimates (MLE). Results of the KS tests are reported in Tab.~\ref{tab2}.
By considering a level of significance $\alpha = 0.01$, we fail to reject the null hypothesis of equivalence of the two distributions in all cases. The Logistic model maximizes the log-likelihood for both $S_5$ and $C_5$.
\begin{table}[ht]
	\centering
	\begin{tabular}{ c | c| c | c c |c|c|c}
			
		& $D$ &  $C$& $p$ & 	& $D$ &  $C$& $p$  \\ 
		%\hline
		$ \mathit{S_5/GM}$ & $0.071$ & $0.079$ & $0.031$  &	$ \mathit{C_5/GM}$ & $0.060$ & $0.079$ & $0.100$  \\
		$ \mathit{S_5/LM3}$& $0.061$ & $0.079$ & $0.089$ &$ \mathit{C_5/LM3}$ & $0.077$ & $0.079$ & $0.015$ \\
		$ \mathit{S_5/LM5}$ & $0.049$ & $0.079$ & $0.267$ &	$ \mathit{C_5/LM5}$& $\mathbf{0.050}$ & $0.079$ & $0.241$ \\
		$ \mathit{S_5/LLM}$& $\mathbf{0.047}$ & $0.079$ & $0.322$ &	$ \mathit{C_5/LLM}$& $0.053$ & $0.079$ & $0.197$ \\
			
	\end{tabular}
	\caption{Results from Kolmogorov-Smirnov test. $D$ is the estimated maximum distance between the two distributions under analysis, $C$ is the corresponding critical value, and $p$ the resulting p-value. Considering a level of significance $\alpha = 0.01$, we fail to reject the null hypothesis of equivalence of the two distributions in all cases.}\label{tab2}
\end{table}

The particular \textit{S}-shaped behavior observed on raw data, and then characterized by growth-model fits, reminds the one observed in the framework of population growth, where after a first stage of huge growth, a saturation level is reached, and population stabilizes. Logistic and Gompertz growth models found several fields of application, ranging from demography and sociology, to biology and ecology \cite{milotti2012interplay, tindall2008modelling}. 
\\
Science and conspiracy communities reach a thresholding value in their sizes growth, as fit results suggest. Those users which are deeply engaged in a community are more likely to become focused on a particular topic, and their increasing involvement into highly specified topics makes them ``isolated'' from the neighboring environment, which in this case is the whole world of knowledge. What is curious is that both conspiracy and science communities show the same size profiles.
\\
To better assess the reliability of model fits results, we further inspect the time evolution of $S_5$ and $C_5$ communities sizes through advanced spectral methodologies extremely useful to uncover the presence of significant oscillatory movements, besides the huge growing trend dominating both communities temporal evolution.
More precisely, we try to identify trends, oscillatory components (both periodic and not-periodic), and background noise in our series to finally reconstruct the embedded true signal, by summing up the contributions of all its significant components. 
We chose non-parametric methods, such as singular-spectrum analysis and similar methodologies\cite{ghil2002advanced,GolyaZhi2013}, in order to analyze our records time evolution by an alternative approach, which is not based on fitting an assumed model to the data, with the final goal in mind to further support model fits results by a completely different method. 
Indeed, the simultaneous and flexible application of more than one spectral tool can assure a quite reliable and robust analysis of temporal dynamics, especially when the signal-to-noise ratio is low, besides dealing with finite sample length. 
Moreover Monte-Carlo SSA (MCSSA)\cite{allensmith1996,groth2015monte} test is applied to assess the significance of the revealed oscillatory modes with respect to both white and red noise background noise null-hypotheses. The reader is referred to Section ~\textit{Methods} for deeper details about the applied methodology.

\begin{figure}[h]
	\centering
	\includegraphics[width=1\textwidth]{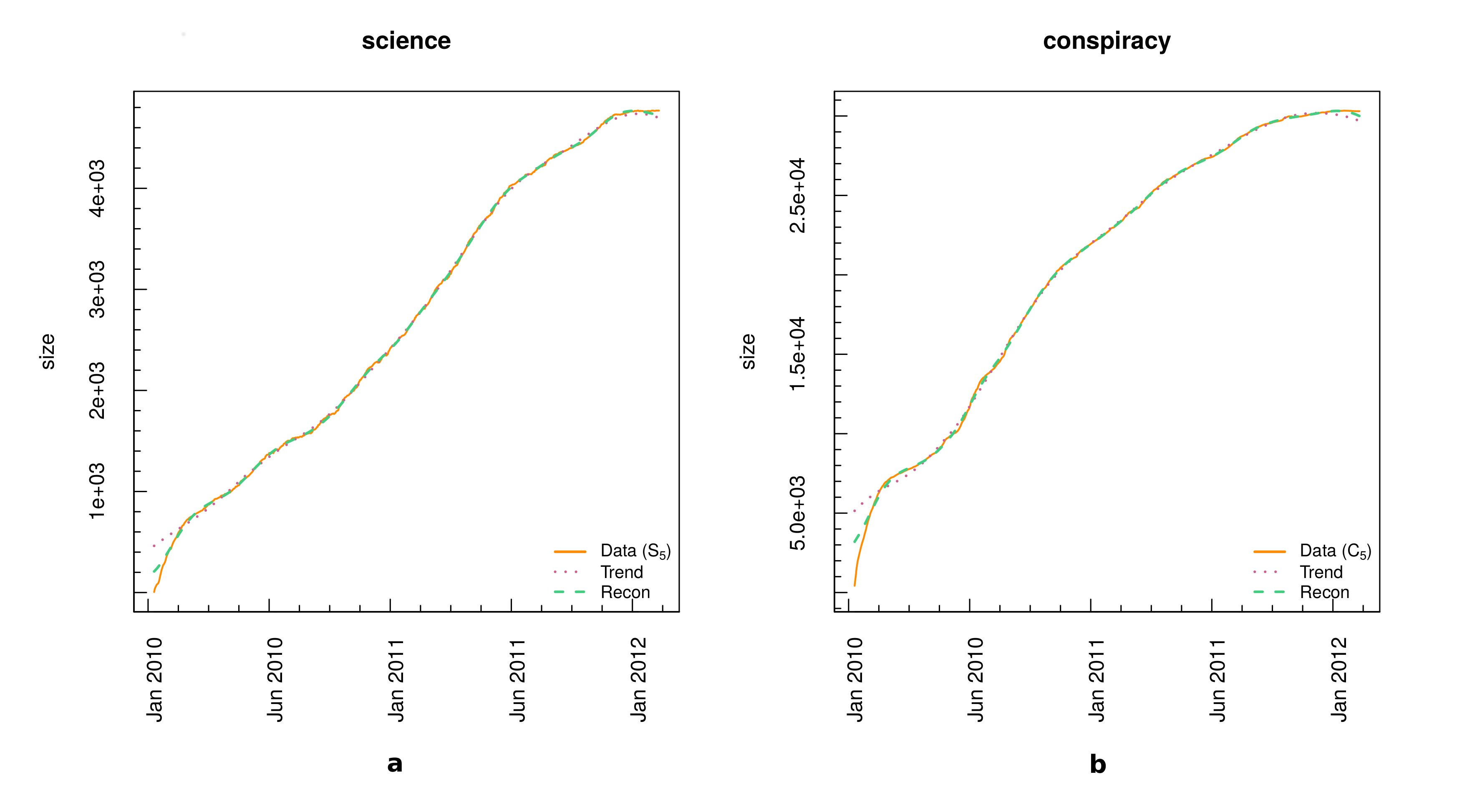}
	\caption{$S_5$ (a) and $C_5$ (b) dominant spectral components. Original series are shown in orange lines, trends in dotted violet lines, and significant signal reconstructions in dashed green lines. A pervasive trend dominates both science and conspiracy communities sizes temporal evolution. The significant reconstructed signals, as well, are led by the trends behaviour.}\label{fig:trends}
\end{figure}

Both conspiracy and science time series behavior turn out to be described by the first two T-PCs (temporal principal components), which in that case correspond to the trend. More in details, the trends capture the $96.16\%$ and the $95.44\%$ of $S_5$ and $C_5$ series total variance, respectively. Besides, we extracted the pure significant reconstructed signals from our series, and we observed that they turned out to be quite similar to trends (exception made for some boundary effects due to the finite-sample length). 
Figure~\ref{fig:trends} shows the trends (dotted violet line) and reconstructed signals (dashed green line) superposed to $S_5$ (panel a) and $C_5$ (panel b) communities size evolution in time (orange lines). Boundary effects are visible, especially at the beginning of the series, but quite negligible. Trends are able to catch both $S_5$ and $C_5$ temporal profile, and they mainly coincide with the reconstructed significant signals, in both cases. 

\begin{figure}[h]
	\centering
	\includegraphics[width=1\textwidth]{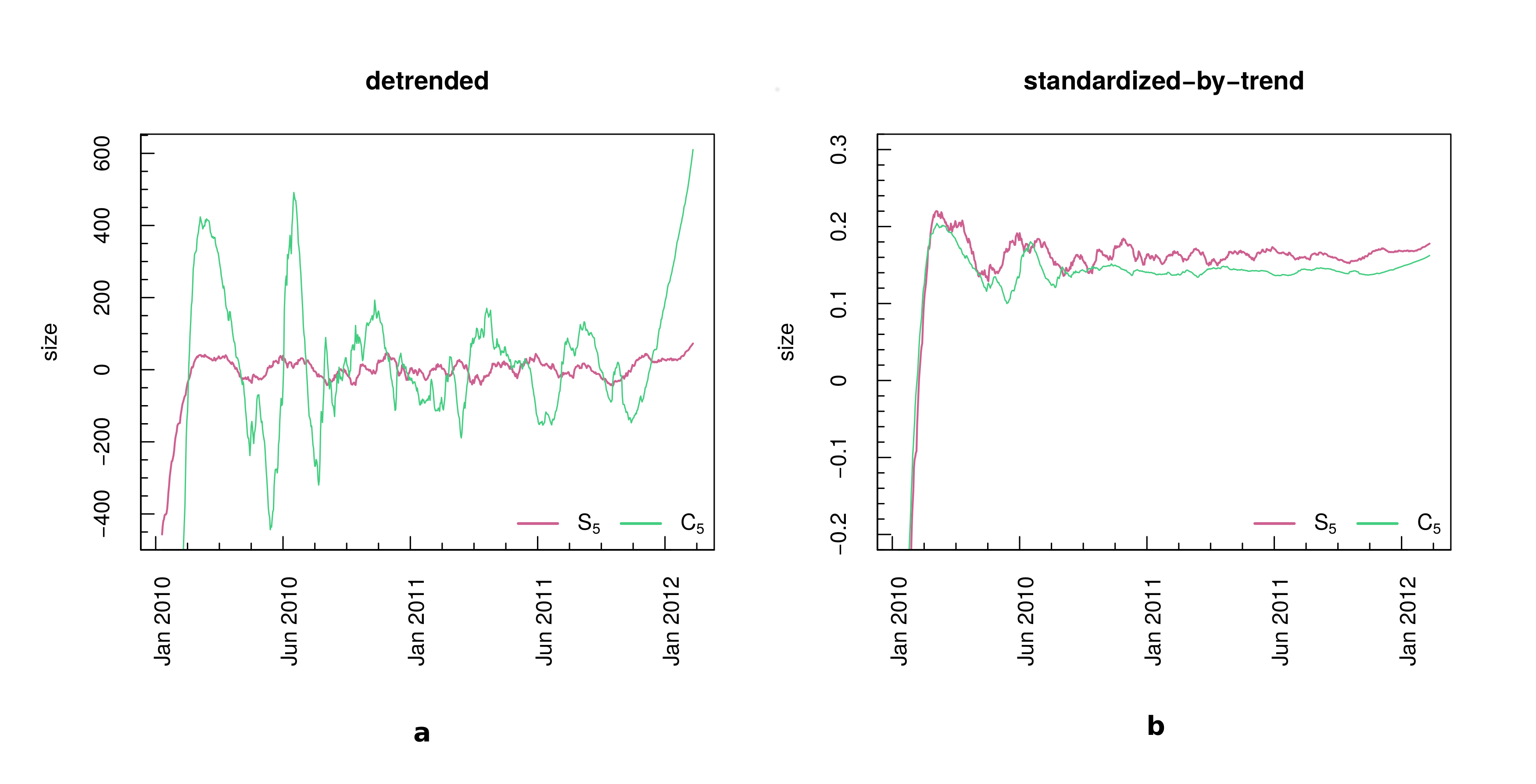}
	\caption{Pre-processing procedure. (a) Detrended $S_5$ (solid pink) and $C_5$ (solid green). (b) Standardized-by-trend $S_5$ (solid pink) and $C_5$ (solid green) residual time series. In panel b, the pre-processed series are standardized to zero mean and unit variance.}\label{fig:detrstdbytrends}
\end{figure}

As a further check, we pre-process data, first by removing the trend, second by standardizing-by-trends the so obtained residual time series.\footnote{Pre-processing is required since the presence of such a pervasive trend reflects in a high peak at zero frequency dominating the shape of power spectrum estimate, and sometimes hiding eventual higher-frequency cycles.} No significant cycle is detected in $S_5$ and $C_5$ series after trend removal.
Figure~\ref{fig:detrstdbytrends} shows $S_5$ and $C_5$ detrended time series (panel a) and $S_5$ and $C_5$ residual time series standardized by their trends (panel b)\footnote{In order to help the visual comparison between $S_5$ and $C_5$, both the pre-processed time series shown in Fig.\ref{fig:detrstdbytrends}b are standardized to zero mean and unit variance.}. The apparent oscillating behavior visible in raw data and in the detrended time series (especially in $C_5$, Fig.~\ref{fig:detrstdbytrends}a) is not connected to significant oscillatory modes, according to Monte-Carlo SSA test. Besides, both communities show a smoother profile after Jan $2011$ (Fig.~\ref{fig:detrstdbytrends}b), corresponding to the range $t > 600$ in Fig.\ref{fig:community_fit}. At that time, both $S_5$ and $C_5$ growth starts to decrease. 
	
We can finally infer that the trends determine the time evolution of our records, only. Thus, we compare the previous described model fits (GM, LM$3$, LM$5$, LLM) to the $S_5$ and $C_5$ trends, only.
No particular difference emerges between science and conspiracy communities in terms of their growth, and the linear correlation between both communities trends and each fitted model turns to be very high for all the cases, preventing us to identify a significantly favorite fit, in agreement with the results previously reported in Tab.~\ref{tab2}. \textit{Pearson} correlation coefficient is computed, since no particular significant cycle emerges from $S_5$ and $C_5$ sizes records spectral analysis, thus reducing the risk of underestimating the presence of an eventual correlation at time-shifted version of the original series.   

Our analysis thus suggests that communities present strong similarities, and that the behavior of users inside each of them is similar. 
Once they have selected their preferred group, users seem to undergo community dynamics, that are similar in both science and conspiracy case, irrespectively of the content. 	
	
\subsection*{Users' Sentiment Analysis}
Now we zoom in at the level of the emotional dynamics of the polarized groups.
We approximate the emotional attitude of users towards one piece of information that they commented by considering the sentiment of the text. We label the sentiment of each comment as: \textit{negative} (-1), \textit{neutral} (0), or \textit{positive} (+1). 
We perform an automatic sentiment classification based on supervised machine learning, refer to Section \textit{Materials and Methods} or to 
\cite{zollo2015emotional} for more details.
	
Our aim is to characterize the emotional behavior of the users as a function of their involvement inside the community. To do this we define three new measures, the \textit{mean user sentiment} ($\sigma_i$), the \textit{mean negative/positive difference of comments} ($\delta_{NP}(i)$), and the \textit{user sentiment polarization} ($\varrho_{\sigma}(i)$) as it follows:
\begin{equation}
\delta_{NP}(i)= \frac{1}{T_i}\sum_{j = 1}^{T_i}(Neg_j(i) - Pos_j(i)),
\end{equation}
where $T_i$ is the number of days in which user $i$ was active, $Neg_j(i)$ the number of $i$'s negative comments in day $j$, $Pos_j(i)$ the number of $i$'s positive comments in day $j$; 
\begin{equation}
\varrho_{\sigma}(i) = \frac{(N_i - 2k_i - h_i)(N_i - h_i)}{N_i^2}, 
\end{equation}
where $N_i, k_i, h_i$ are respectively the number of all, negative, and neutral comments left by user $i$, while $l_i = N_i - k_i - h_i$ is the number of the positive ones.
Note that $\varrho_{\sigma}(i)\in[-1, 1]$ and that it is equal to $0$ if and only if $l_i  = k_i$ or $h_i = N_i$, it is equal to $1$ if and only if $k_i  = N_i$, and it is 
equal to $-1$ if and only if $l_i  = N_i$. While $\sigma_i$ is simply defined as the mean of the sentiment of all comments left by user $i$.
	
\begin{figure}[h]
	\centering
	\includegraphics[width=1\textwidth]{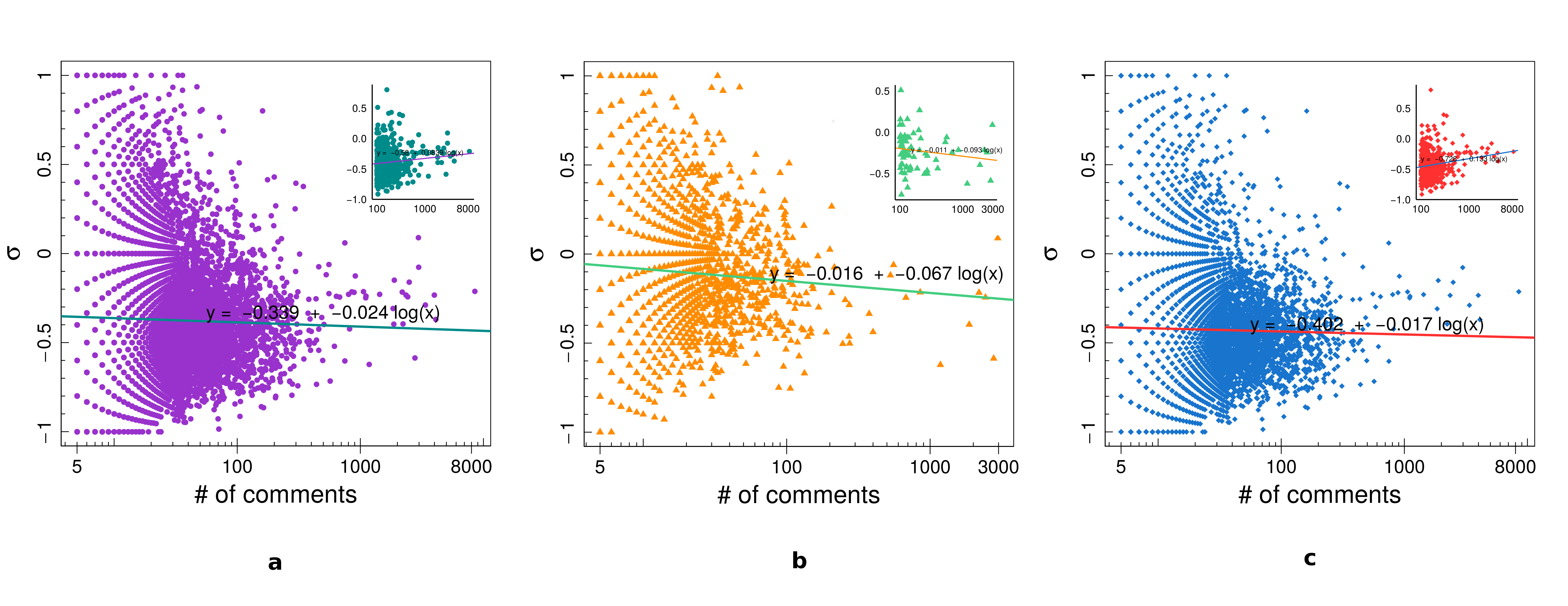}
	\caption{Mean final sentiment $\sigma_i$ of all users (a), science users (b), and conspiracy users (c), as a function of the user engagement i.e., the total number of comments left by each user. In the insets we report, for each of the three categories, the value of $\sigma_i$ as a function of the number of comments for those users with at least 100 comments. We regressed the mean user sentiment $\sigma_i$ w.r.t. the logarithm of the number of comments.}\label{fig:mean_final_sent}
\end{figure}

Figure~\ref{fig:mean_final_sent} shows the average sentiment $\sigma_i$ for all users (panel a), science users (panel b), and conspiracy users (panel c), as a function of the user engagement -- i.e., the total number of comments left by each user. 
In the insets we report, for each of the three categories, the value of $\sigma_i$ as a function of the number of comments for the most active users, i.e. those users with at least 100 comments. 
We then regress the mean user sentiment $\sigma_i$ w.r.t. the logarithm of the number of comments. 
We notice that $\sigma_i$ becomes more negative as the number of comments increases, in all cases. 
	
\begin{figure}[h]
	\centering
	\includegraphics[width=1\textwidth]{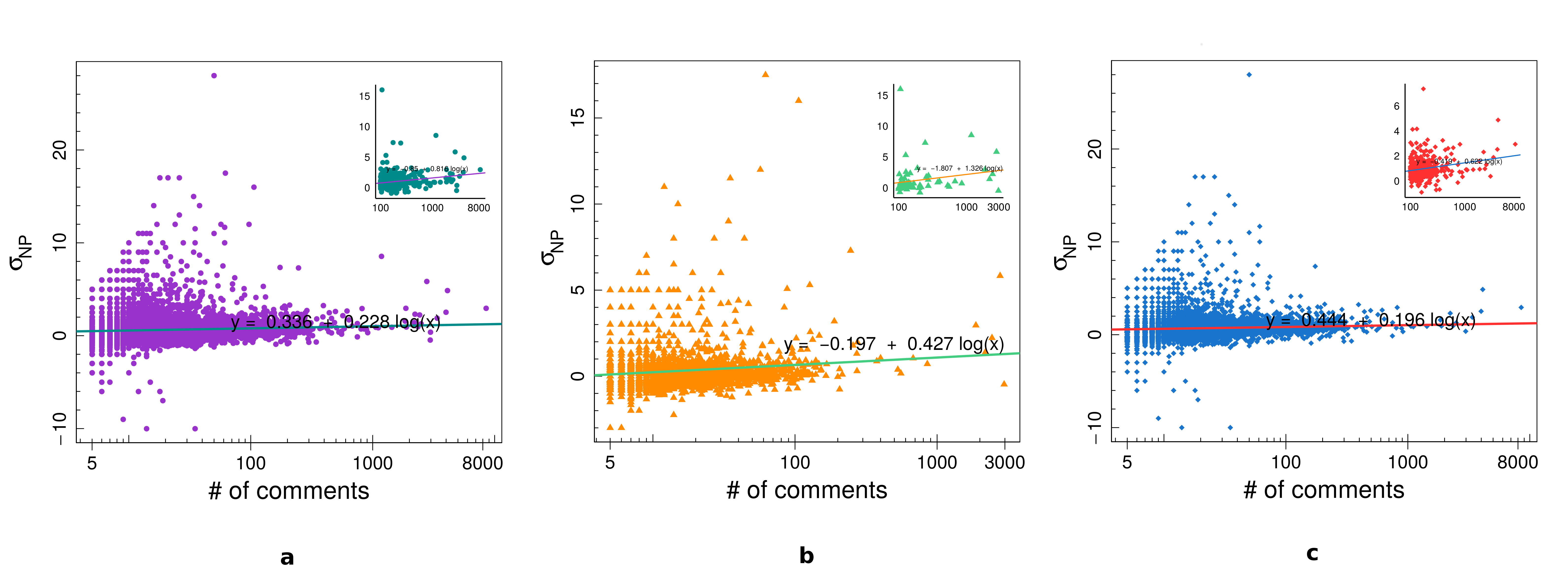}  	
	\caption{Mean negative/positive difference $\delta_{NP}(i)$ of all users (a), science users (b), and conspiracy users (c), as a function of the user engagement. In the insets we report, for each of the three categories, the value of $\delta_{NP}(i)$ as a function of the number of comments for those users with at least 100 comments. We regressed the mean negative/positive difference  $\delta_{NP}(i)$ w.r.t. the logarithm of the number of comments.}\label{fig:mean_np_diff}
\end{figure}

Figure~\ref{fig:mean_np_diff} shows the mean negative/positive difference of comments $\delta_{NP}(i)$ of all users (panel a), science users (panel b), and conspiracy users (panel c), as a function of the user engagement. In the insets we report, for each of the three categories, the value of $\delta_{NP}(i)$ as a function of the number of comments for those users with at least 100 comments. We regressed the mean negative/positive difference $\delta_{NP}(i)$ w.r.t. the logarithm of the number of comments. $\delta_{NP}(i)$  is a measure of the mean negative shift from a situation of neutral equilibrium for which either the user has only neutral comments or he/she has the same number of positive and negative comments. A positive value of $\delta_{NP}(i)$ indicates that the user tends to have, on average, more negative than positive comments. From Fig. \ref{fig:mean_np_diff} we notice that $\delta_{NP}(i)$ tends to increase when the number of comments increases in all cases, underlining the fact that, on average, more active users tend to show a faster shift towards the negativity than less active ones.
The rate of this increment in the negativity is higher for users with more than 100 comments and it is also higher for science users w.r.t conspiracy ones.
	
\begin{figure}[h]
	\centering
	\includegraphics[width=1\textwidth]{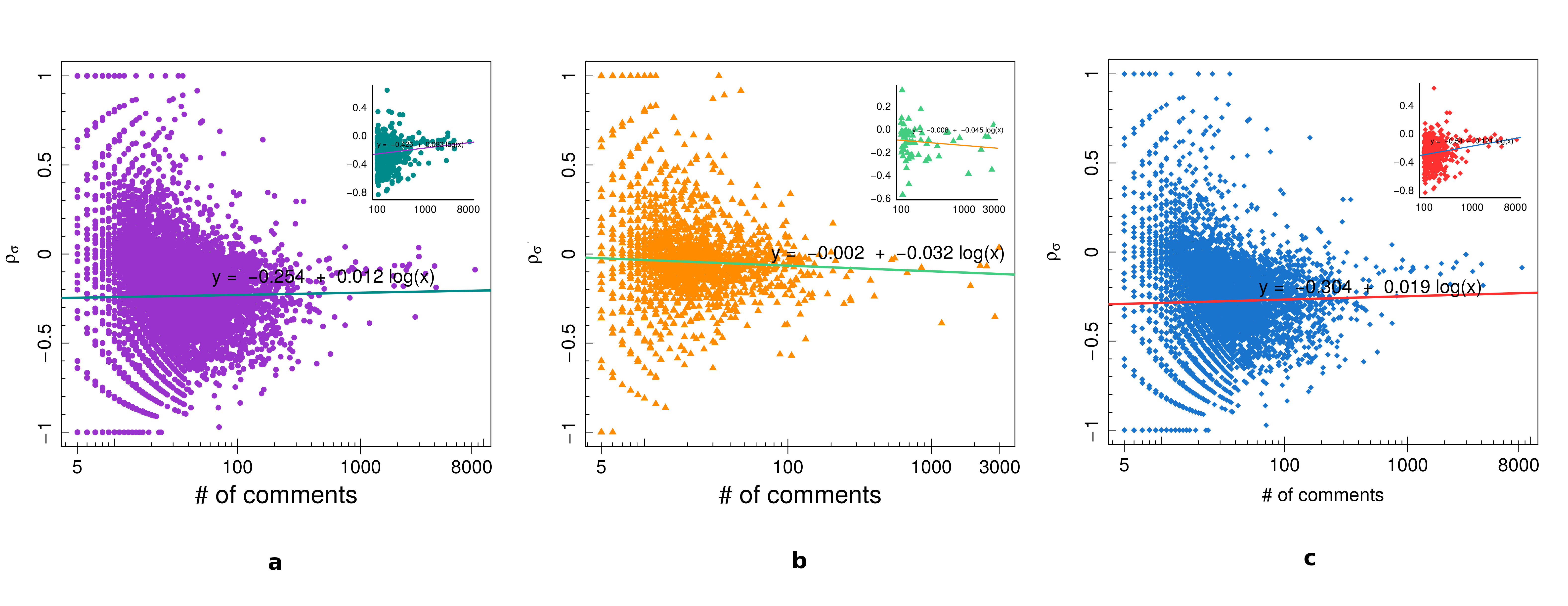}
	\caption{User's sentiment polarization  $\varrho_{\sigma}(i)$ of all users (a), science users (b), and conspiracy users (c), as a function of the user engagement. In the insets we report, for each of the three categories, the value of $\varrho_{\sigma}(i)$ as a function of the number of comments for those users with at least 100 comments. We regressed the user sentiment polarization $\varrho_{\sigma}(i)$ w.r.t. the logarithm of the number of comments.}\label{fig:user_sent_pol}
\end{figure}

Figure~\ref{fig:user_sent_pol} displays the user sentiment polarization $\varrho_{\sigma}(i)$ of all users (panel a), science users (panel b), and conspiracy users (panel b), as a function of the user engagement. In the insets we show, for each of the three categories, the value of $\varrho_{\sigma}(i)$ as a function of the number of comments for those users with at least 100 comments. We regressed the user sentiment polarization $\varrho_{\sigma}(i)$ w.r.t. the logarithm of the number of comments.
The user sentiment polarization $\varrho_{\sigma}(i)$ ranges in $[-1, 1]$, and it is equal to 0 either if all comments are neutral or if there is the same number of negative and positive comments, while it tends to 1 (resp. -1) when $l_i \gg k_i$ and $h_i$ is small enough, i.e., when the number of positive comments is much bigger than the number of negative ones, (resp. $k_i \gg l_i$ and $h_i$ is small enough, i.e., when the number of negative comments is much bigger than the number of positive ones).
Science users show an higher value of $\varrho_{\sigma}(i)$, however conspiracy users with at least 100 total comments tend to increase it w.r.t. science ones.	
	
The engagement within the echo chamber affects users emotional dynamics.
The more a user is active, the higher the tendency to express negative emotion when commenting. This feature holds for both users categories. 
Moreover, for both categories we observe that, on average, more active users show a faster shift towards the negativity than less active ones. The rate of this increment in the negativity is higher for users with more than 100 comments and it is also higher for science users w.r.t conspiracy ones. In terms of the users' sentiment polarization we observe some differences between the two categories: its value is generally higher for science users, however very active science users tend to decrease their sentiment polarization with the increasing of the activity, while on the contrary conspiracy ones tend to increase it.
	
\subsection*{Evolution of the Sentiment inside the Communities}
We now focus on the collective sentiment of the two communities, rather than the single user's one. Similarly to the single user case we define the \textit{community negative/positive difference} of comments ($\delta^C_{NP}$) and the \textit{mean community sentiment polarization} ($\varrho_{\sigma}^C$) as follows:
\begin{equation}
\delta^C_{NP} = \frac{1}{M_C}\left(\frac{1}{T}\sum_{j = 1}^T(Neg^C_j - Pos^C_j)\right),
\end{equation}
where $T$ is the number of days of observations, $Neg^C_j$ the number of negative comments from users belonging to community $C$ during day $j$, $Pos^C_j$ the number of positive comments from users belonging to community $C$ during day $j$, $M_C$ is the maximum daily activity of community $C$, and $C \in\left\{ Science, Conspiracy \right\}$, while
\begin{equation}
\varrho_{\sigma}^C = \frac{(N_C - 2k_C - h_C)(N_C - h_C)}{N_C^2}, 
\end{equation}
where $N_C, k_C, h_C$ are respectively the number of all, negative, and neutral comments left by users of community $C$, while $l_C = N_C - k_C - h_C$ is the number of positive ones.
Note that $\varrho_{\sigma}^C \in[-1, 1]$.
	
\begin{figure}[h]
	\centering
	\includegraphics[width=0.6\textwidth]{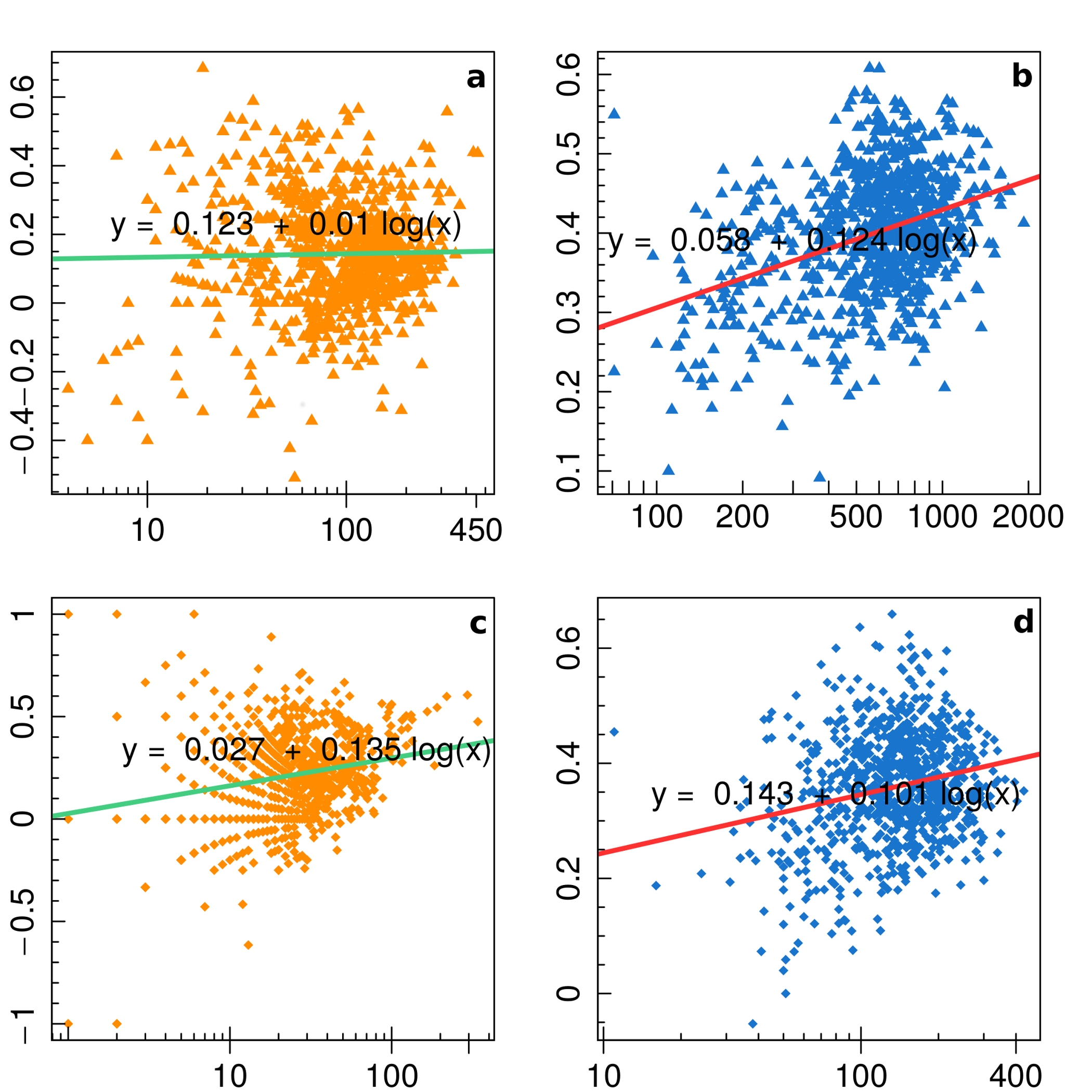}
	\caption{Community negative/positive difference of comments  $\delta^C_{NP}$ as a function of the daily community activity for science users (a, c) and conspiracy users (b, d). The top figures (a, b) show the values of $\delta^C_{NP}$ considering all users in the communities, while the bottom ones (c, d) only consider those users with at least 100 comments. We regressed the community negative/positive difference of comments $\delta^C_{NP}$ (y-axes) w.r.t. the logarithm of the number of comments inside the community at a given time (x-axes).}\label{fig:community_nd}
\end{figure}

Figure~\ref{fig:community_nd} displays the community negative/positive difference of comments $\delta^C_{NP}$ as a function of the daily community activity for science users (left: panels a, c) and conspiracy users (right: panels b, d). The top figures (panels a, b) show the values considering all users in the communities, while the bottom ones (panels c, d) only consider those users with at least 100 comments. We regressed the community negative/positive difference of comments $\delta^C_{NP}$ (y-axes) w.r.t. the logarithm of the number of comments inside the community at a given time (x-axes). For both communities $\delta^C_{NP}$ tend to increase, while science one shows an higher increasing rate for the most active case, conspiracy one shows an higher increasing rate for the general case.
	
\begin{figure}[h]
	\centering
	\includegraphics[width=0.6\textwidth]{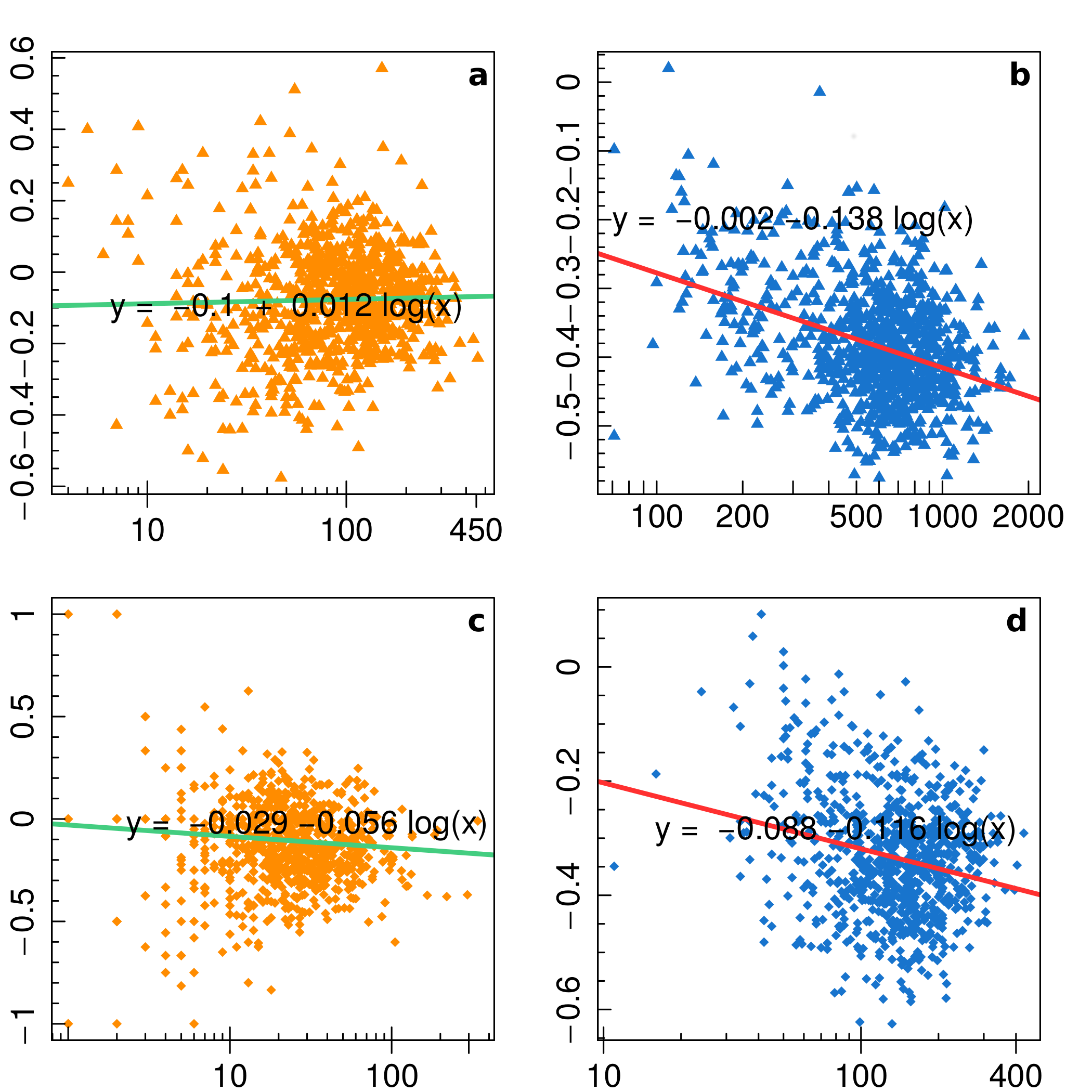}
	\caption{Mean community sentiment polarization $\varrho^C_{\sigma}$ as a function of the daily community activity for science users (left) and conspiracy users (right). The top figures show the values of $\varrho^C_{\sigma}$  considering all users in the communities, while the bottom ones only consider those users with at least 100 comments. We regressed the community negative/positive difference of comments $\varrho^C_{\sigma}$ (y-axes)  w.r.t. the logarithm of the number of comments inside the community at a given time (x-axis).}\label{fig:community_mbip}
\end{figure}

Figure \ref{fig:community_mbip} shows the mean community sentiment polarization $\varrho^C_{\sigma}$ as a function of the daily community activity for science users (left: panels a, c) and conspiracy users (right: panels b, d). The top figures display the values considering all users in the communities (panels a, b), while the bottom ones only consider those users with at least 100 comments (panels c, d). As for Fig.~\ref{fig:community_nd}, we regressed the mean community sentiment polarization $\varrho^C_{\sigma}$ w.r.t. the logarithm of the number of comments inside the community at a given time. For the conspiracy community we notice a decrement in the value of $\varrho^C_{\sigma}$ as the number of comments increases, moreover this decrement is higher for most active users. Science community instead shows a decrement in the value of $\varrho^C_{\sigma}$ for the case of most active users and a slight increment for the general case. 
	
Also the community sentiment behavior is affected by the cumulative users' activity (in terms of comments). When either community is more active, the shift towards negative comments is larger. A difference between the two echo chambers comes upon if we restrict our attention only to the most active users, i.e. those with at least 100 comments. In this last case, science users show a higher rate of increment than conspiracy ones, contrary to the general case. Differently from the single user case, the community sentiment polarization shows a deep decrement with higher activity in the conspiracy community, the process is slower for science community, when we consider only most active users, and even reversed in the general case. 
	
\section*{Conclusions}
The Facebook environment is particularly suited for the emergence of polarized communities, or echo chambers. The activity inside such echo chambers is limited to only one type of content.
In this work, we characterize the behavior of users inside the echo chamber and the structural evolution of the community accounting for both users activity and the sentiment they express. 
	
We first study the evolution of the size of the two communities by fitting daily resolution data with three growth models, i.e. the Gompertz model, the Logistic model, and the Log-logistic model, and we observe that both communities evolve in a similar way and the behavior of users is similar irrespectively of the difference in contents: after a first phase of rapid growth, approximately exponential, both the communities sizes undergo a more gradual growth, till a thresholding value is reached. The lack of communication with the environment can be supposed to associate with the users extreme focusing on a precise topic.
	
Then we notice that both the users' and the communities' emotional behavior is affected by the users' involvement inside the echo chamber. To an higher involvement corresponds a more negative approach.  Moreover, for both categories we observe that, on average, more active users show a faster shift towards the negativity than less active ones. The rate of this increment is higher for users with more than 100 comments and higher for science users w.r.t conspiracy ones. The community sentiment polarization shows a deeper decrement with higher activity in the conspiracy community, while the process is slower for science community, when we consider only most active users, and even reversed in the general case.
	
\section*{Methods}
\subsection*{Ethics Statement}
The data collection process has been carried out using the Facebook Graph application program interface (API) \cite{API}, which is publicly available. For the 	analysis (according to the specification settings of the API) we only used publicly available data (thus users with privacy restrictions are not included in the dataset). The pages from which we download data are public Facebook entities and can be accessed by anyone. User content contributing to these pages is also public unless the user’s privacy settings specify otherwise, and in that case it is not available to us.	
\subsection*{Data Collection and Description} 
Using the approach described in \cite{bessi2015science}, with the support of diverse Facebook groups very active in the debunking of misinformation (Protesi di Complotto, Che vuol dire reale, La menzogna diventa verit\`{a} e passa alla storia), we identified two main categories of pages: \textit{conspiracy theories}, i.e., pages promoting contents neglected by main stream media, and \textit{science information}, i.e., pages diffusing scientific news and research advances for which it is easy to check the sources. Starting from this basic differentiation, we categorized Facebook pages according to their contents and their self description.
The resulting dataset is composed of 73 public Italian Facebook pages, 34 of which were diffusing scientific information and 39 conspiracy theories, and covers a timespan of 5 years, from 2010 to 2014. Table \ref{tab0} summarizes the details of our data collection.	
\begin{table}[ht]
		
 	\centering
	\begin{tabular}{l | c c c }
	
		& \textit{Total} & \textit{Science} & \textit{Conspiracy} \\ [0.5ex] 
	 	\hline
		\textit{Pages} & 73& 34  & 39 \\ 
		\textit{Posts} & 271,296& 62,705 & 208,591\\
		\textit{Likes} & 9,164,781   & 2,505,399 & 6,659,382\\[1ex]
		\textit{Comments} & 1,017,509  & 180,918 & 836,591 \\[1ex]
		
	\end{tabular}
	\caption{Dataset description.}\label{tab0} 
\end{table}
	
\subsection*{Growth Models}
The \textbf{Gompertz Growth Model} is often used to model growth phenomena which are typically characterized by an asymptotic behaviour rather than by a linear increase. In that sense, a Gompertz function has to be intended as a special case of the most general logistic function, and it is nowadays applied in various research fields, such as biology, ecology, economics, marketing, and medicine. In oncology, in particular, the Gompertz sigmoid function has been used to model tumor growth \cite{ferrante2000theprob, Jukic2004least}, which are interpreted as an expansion of cellular populations developing in a confined space, where the availability of nutrients is limited in a certain sense. As a consequence, the model considers two parameters, a first one, $a$, for the tumor intrinsic growth related to the mitosis rate and a second one, $b$, for the growth deceleration, due to the antiangiogenic processes.  Let $x(t)$ be the size of the tumor at time $t$, then we have:
$$
\frac{dx(t)}{dt} = a x(t) - b x(t) \ln x(t).
$$
For a given initial condition $x(0) = x_0$, and known parameters $a$ and $b$, the solution \cite{ferrante2000theprob} is:
\begin{equation}\label{GM}
x(t) = e^{a/b +(\ln(x_0) - a/b)e^{-bt}}.
\end{equation}
	
The most general \textbf{Logistic Growth Model} is defined as:
\begin{equation}\label{LM}
x(t) = c + \frac{d-c}{(1+ e^{b(t - f)})^g}.
\end{equation}
In that case, the first stage of the growth is approximately exponential, then the growth rate decreases till an asymptotic value is reached. That right-hand asymptote is reached less gradually than the left-hand one compared to the behaviour of the Gompertz function. We used two variants of the Logistic model to fit our data: $L3$ that considers only parameters $(b, d, f)$ in (\ref{LM}), and $L5$ that is exactly (\ref{LM}).

Finally, the \textbf{Log-Logistic Growth Model} is defined as:
\begin{equation}\label{LLM}
x(t) =  c + \frac{d-c}{(1+e^{b(\ln t -\ln f)})^g}.
\end{equation}

\subsection*{Nonlinear Least Square Fitting and Goodness of Fit}
We use the \textit{Nonlinear Least Squares} (NLS) \cite{Jukic2004least} to estimate the parameters of the various models while fitting them with our data. Consider a set of $n$ observations $(t_1, x_1),\ldots, (t_n, x_n)$ and a model function depending on $m$ parameters $y = f(x,\boldsymbol \beta)$, where $\boldsymbol \beta = (\beta_1, \ldots, \beta_m)$ and $n\geq m$. We want to find the vector $\boldsymbol \beta$ that minimizes the sum of squares:
$$
S= \sum_{i =1}^n r_i^2,
$$
where the residuals errors $r_i$ are given by
$$
r_i= y_i - f(x_i, \boldsymbol \beta),
$$
for $i=1, 2,\dots, m$.

We tested the goodness of our fit by means of the \textit{Kolmogorov-Smirnov Test}. 
	
\subsection*{Advanced spectral analysis and trend extraction procedure}
\textbf{Singular-spectrum analysis (SSA)} is a not-conventional spectral analysis method which provides insight into the unknown and/or partially known dynamics of a dynamical system \cite{ghil2002advanced, GolyaZhi2013}. More in detail, SSA aims at decomposing the signal as a linear combination of variability modes, which are data-adaptive functions of time. Thus, with respect to more traditional spectral approaches such as the classical Fourier decomposition, SSA doesn't ground on variability modes which have to be necessarily harmonic components. As a consequence, SSA provides a powerful de-noising filter, to identify the different components of the analyzed signal, such as trends, oscillatory patterns, harmonic and/or anharmonic oscillations, quasi-periodic phenomena, without making any assumption about the underlying generating model of the  observed signal \cite{alessio2016digital}. Moreover, SSA doesn't require the assumption of any particular stationarity or ergodicity conditions.
	
In order to distinguish between significant signal and random fluctuations (i.e. background noise), \textbf{Monte-Carlo SSA (MCSSA)} is applied. MCSSA grounds on a particular Monte Carlo approach to the signal-to-noise separation issue, suited to overcome the limitations of classical signal extraction procedure, i.e. the identification of simply a gap in the eigenvalues spectrum \cite{allensmith1996}. Recent fine-tunings of the method have been proposed to further improve results robustness and reliability in short time series \cite{groth2015monte}. In the present work, MCSSA is applied to $S_5$ and $C_5$ time series, to establish whether our time series are linearly distinguishable from the linear stochastic processes, usually considered as noise. Both white and red noise null-hypotheses are taken into account, since the choice of the most suitable kind of noise in social sciences, when dealing with advanced spectral methodologies, is still under debate. 
	
\subsection*{Sentiment Classification}
The sentiment classification is carried out as in \cite{zollo2015emotional} and refers to the same dataset. We consider three values for the sentiment of each comment: \textit{negative} (-1), \textit{neutral} (0), and \textit{positive} (+1). We perform an automatic sentiment classification based on supervised machine learning that consists of the following four steps: (i) a sample of texts is manually annotated with sentiment (in our case 20K randomly selected comments are manually annotated by 22 native Italian speakers), (ii) the labeled set is used to train and tune a classifier, (iii) the classifier is evaluated on an independent test set or by cross-validation, and (iv) the classifier is applied to the whole set of texts. For more details on the classifier or on its performance refer to \cite{zollo2015emotional}.

%\bibliography{refs}

\section*{Acknowledgements}
Funding for this work was provided by EU FET project MULTIPLEX nr. 317532, SIMPOL nr. 610704, DOLFINS nr. 640772, SOBIGDATA nr. 654024. The funders had no role in study design, data collection and analysis, decision to publish, or preparation of the manuscript. We want to thank Geoff Hall and “Skepti Forum” for providing fundamental support in defining the atlas of conspiracy news sources in the US Facebook.

\section*{Additional information}

\textbf{Competing financial interests} The authors declare no competing financial interests.

\end{document}